\documentclass{statsoc}

\usepackage{times}
\usepackage{bm}
\usepackage{amssymb}
\usepackage{graphicx}
\usepackage{natbib}
\usepackage{amsmath}
\usepackage[plain,noend]{algorithm2e}
\usepackage{breqn}
\usepackage{xcolor}
\usepackage{multirow}
\usepackage{enumitem}
\usepackage[most]{tcolorbox}

\usepackage[a4paper]{geometry}
\usepackage[textwidth=8em,textsize=small]{todonotes}
\usepackage{hyperref} 
\usepackage{comment}
\usepackage{blindtext}
\usepackage{float}
\usepackage{tikz}
\usetikzlibrary{arrows.meta, positioning, shapes}

\usepackage[nomarkers]{endfloat} 

\usepackage{changepage}

\title[]{Estimating Global HIV Prevalence in Key Populations: A Cross-Population Hierarchical Modeling Approach}
\author[Zhang et. al.]{Jiahao Zhang}
\address{Department of Statistics, Pennsylvania State University, State College, PA 16802 U.S.A.}
\email{jzz5603@psu.edu}
\author[Zhang et. al.]{Keith Sabin}
\address{Strategic Information and Evaluation Division, UNAIDS, 20 Avenue Appia, 1211 Geneva 27, Switzerland}
\email{SabinK@unaids.org}
\author[Zhang et. al.]{Le Bao}
\address{Department of Statistics, Pennsylvania State University, State College, PA 16802 U.S.A.}
\email{lebao@psu.edu}
\begin{document}
\begin{abstract}
Key populations at high risk of HIV infection are critical for understanding and monitoring HIV epidemics, but global estimation is hampered by sparse, uneven data. We analyze data from 199 countries for female sex workers (FSW), men who have sex with men (MSM), and people who inject drugs (PWID) over 2011–2021, and introduce a cross-population hierarchical model that borrows strength across countries, years, and populations. The model combines region- and population-specific means with country random effects, temporal dependence, and cross-population correlations in a Gaussian Markov random-field formulation on the log-prevalence scale. In 5-fold cross-validation, the approach outperforms a regional-median baseline and reduced variants (65\% reduction in cross-validated MSE) with well-calibrated posterior predictive coverage (93\%). We map the 2021 prevalence and quantify the change between 2011 and 2021 using posterior prevalence ratios to identify countries with substantial increases or decreases. The framework yields globally comparable and uncertainty-quantified country-by-year prevalence estimates, enhancing evidence for resource allocation and targeted interventions for marginalized populations where routine data are limited.

\end{abstract}
\keywords{HIV Epidemic, Key Population, Sparse Data, Hierarchical Model, Small Area Estimation}

\section{Introduction}
\label{sec:int}
HIV/AIDS remains a global health threat, since the incidence for 2020 was almost three times higher than the interim target of the Sustainable Development Goal of fewer than 500,000 new infections \citep{abd2017global, bekker2023hiv}. Despite significant progress in HIV prevention, diagnosis, and treatment, these advances remain unevenly distributed across populations and regions. Some key populations are still at higher risk for HIV, according to sexual practices, occupations, and substance use \citep{UNAIDS2020a, UNAIDS2020b}. These groups not only experience a disproportionately higher prevalence of HIV, but also face considerable barriers to accessing healthcare and preventive services \citep{hakim2018gaps, garnett2021reductions}. The updated global AIDS response highlights ending inequalities with a particular focus on the HIV epidemic among key populations. To facilitate evidence-based health policy, there is an urgent need for detailed and reliable estimates of HIV epidemics among key populations at both the national and global levels \citep{sullivan2021epidemiology}.

However, estimating HIV prevalence among key populations presents substantial challenges. These groups, including female sex workers (FSW), men who have sex with men (MSM), and people who inject drugs (PWID), often face legal, social and structural barriers that hinder their inclusion in public health surveillance \citep{decker2022systematic}. As a result, data collection is frequently incomplete, geographically uneven, and especially limited in resource-constrained settings, preventing effective monitoring and intervention planning at both the national and global levels \citep{stevens2024population}. Previous approaches to addressing these limitations have included methods such as median imputation, hierarchical Bayesian models, and machine learning techniques \citep{williams2020trends, bao2024dynamic, marcus2020artificial}. These methods have made progress in exploiting existing data. However, they primarily focus on single country or regional contexts and do not fully account for the complexities of global HIV modeling. In particular, challenges such as data sparsity, heterogeneity, and structural dependencies across populations and locations remain difficult to address. Currently, UNAIDS summarizes recent HIV prevalence estimates using a simple regional median approach, typically aggregating the most recent data from the past five years. While practical, this method prioritizes convenience over statistical rigor and does not fully leverage the available data structure across populations and locations \citep{UNAIDS_Atlas}.

In this paper, we propose a cross-population hierarchical model that integrates small area estimation techniques and Gaussian Markov random fields (GMRF) to predict HIV prevalence rates for female sex workers (FSW), men who have sex with men (MSM) and people who inject drugs (PWID) in 199 countries between 2011 and 2021. Our model effectively borrows strength across multiple dimensions -- spatial, temporal, and cross-population structures -- allowing for improved estimation in regions with limited data. This approach addresses data sparsity while capturing key dependencies within countries, over time, and among key populations. To our knowledge, this is the first framework to jointly model multiple key populations across all countries, incorporating cross-population information to bridge gaps that previous methods have not addressed. By comparing our approach to traditional median imputation methods and other reduced models, we demonstrate its superior performance in terms of both accuracy and interpretability.

The remainder of the paper is organized as follows. Section~\ref{sec:data} introduces the motivating dataset and the key challenges in global HIV surveillance. Section \ref{sec:Method} presents our proposed hierarchical model, outlining its structure and statistical formulation. For completeness, we summarize the conditional formulas for unobserved values and Gibbs sampling algorithm used to reconstruct observed posterior samples. Section~\ref{sec:result} presents the empirical results. We evaluate out-of-sample performance via cross-validation, benchmark against a regional-median baseline, and report country-level estimates. We also summarize cross-population correlations with 95\% credible intervals, include a component-exclusion analysis, and quantify changes from 2011--2021 using posterior prevalence ratios. Finally, Section~\ref{sec:Conclusion} summarizes the findings and discusses implications for surveillance and public health policy.

\section{Global HIV Key Population Data (2011 -- 2021)}
\label{sec:data}

We used the Global HIV Key Population dataset from UNAIDS, covering the years 2011 to 2021. This dataset includes HIV prevalence estimates among key affected populations, including female sex workers (FSW), men who have sex with men (MSM), and people who inject drugs (PWID) in 199 countries. To monitor the HIV epidemic within these groups, data have been collected through a combination of biological and behavioral surveillance surveys, enabling robust comparisons across regions and over time, supporting the calibration of epidemic models and informing the strategic allocation of HIV prevention and treatment resources. Detailed documentation of data collection methodologies, survey instruments, and estimation procedures is available at \url{https://kpatlas.unaids.org/}. Access to the dataset can be requested via the UNAIDS data portal at \url{https://aidsinfo.unaids.org/}.

\begin{figure}[htbp!]
    \centering
    \includegraphics[width=0.8\textwidth]{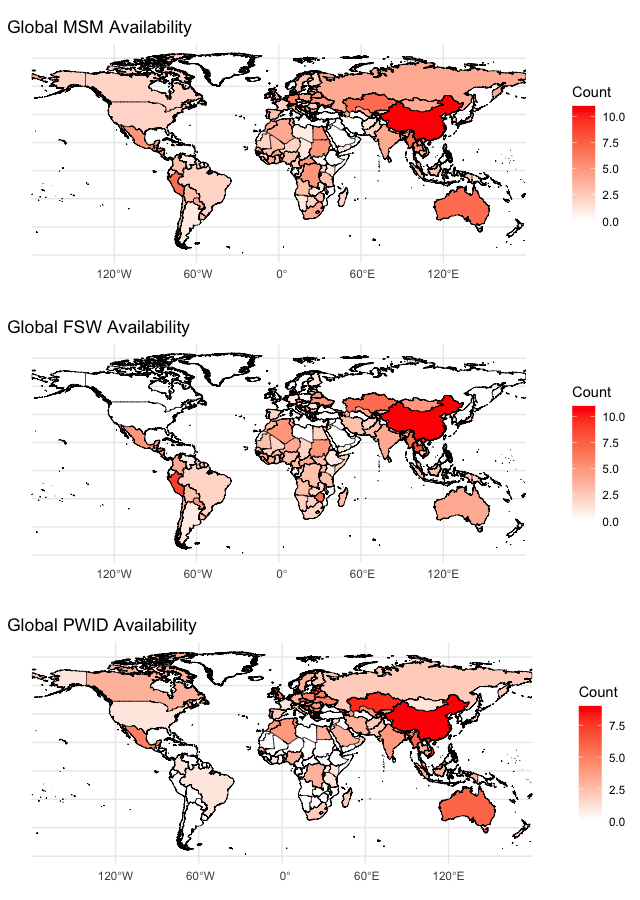}
    \caption{Global data availability for three key populations—men who have sex with men (MSM), female sex workers (FSW), and people who inject drugs (PWID). The color bar indicates the number of available data points per country from 2011 to 2021.}
    \label{fig:Data_dist}
\end{figure}

Although the dataset spans a large number of countries, the availability of prevalence data varies significantly across key populations and regions. Figure \ref{fig:Data_dist} provides a visual representation of data availability for the three key populations from 2011 to 2021. Many countries have very few or no data points, particularly for MSM and PWID. Specifically, for MSM, 23 countries reported five or more data points, 125 countries reported between one and five data points, and 49 countries reported no data. For PWID, 20 countries reported five or more data points, 74 countries reported between one and five data points, and 103 countries reported no data. For FSW, 25 countries reported five or more data points, 108 countries reported between one and five data points, and 63 countries reported no data. This stark data sparsity underscores the challenges inherent in global HIV surveillance and highlights the critical need for methodological approaches to effectively `borrow strength' from related populations and regions and ensure reliable estimation and comparison.


\section{Methodology}
\label{sec:Method}
We introduce the within-country, temporal and cross-population dependencies for efficient estimation of HIV prevalence across country-year-population units with sparse datasets, as outlined in Section \ref{sec:data}. Gaussian Markov Random Fields (GMRF) leverage the sparsity in the precision matrix to encode localized dependencies, enabling scalable inference \citep{rue2005gaussian, lindgren2011explicit, franco2022variance}. We find that the sparse precision matrix facilitates efficient computation, and the Gaussian assumption for log-transformed prevalence rates aligns well with observed data. The section proceeds by outlining the GMRF model, priors, and computational framework. The model evaluation and comparison methods are discussed in Section \ref{sec:result}.

\subsection{Assumptions and Model Components}
Let \( \boldsymbol{Y} \) denote the vector of log transformed HIV prevalence values, where each entry corresponds to a specific country, year, and key population. We write each element as \( Y^{(i)}_{k,\,t} \), where \( i \) indexes countries, \( k \in \{\text{MSM}, \text{FSW}, \text{PWID}\} \) denotes key population, and \( t \in \{2011, \ldots, 2021\} \) indicates year. We assume \( \boldsymbol{Y} \) follows a multivariate normal distribution:
\begin{equation}
\label{eq:GMRF}
\boldsymbol{Y} \sim \mathcal{N}(\boldsymbol{\mu},\, \Sigma),
\end{equation}
where \( \boldsymbol{\mu} \) is the mean vector and \( \Sigma \) is a structured covariance matrix encoding dependencies across time, populations, and countries. 

For computational efficiency and interpretability, we model temporal and cross-population dependencies through a precision matrix \( Q \), and incorporate country-level random effects separately in \( \Sigma \). Details of each component are provided in the following sections.

\begin{figure}[htbp!]
    \hspace{2cm}
    \includegraphics[width=0.8\textwidth]{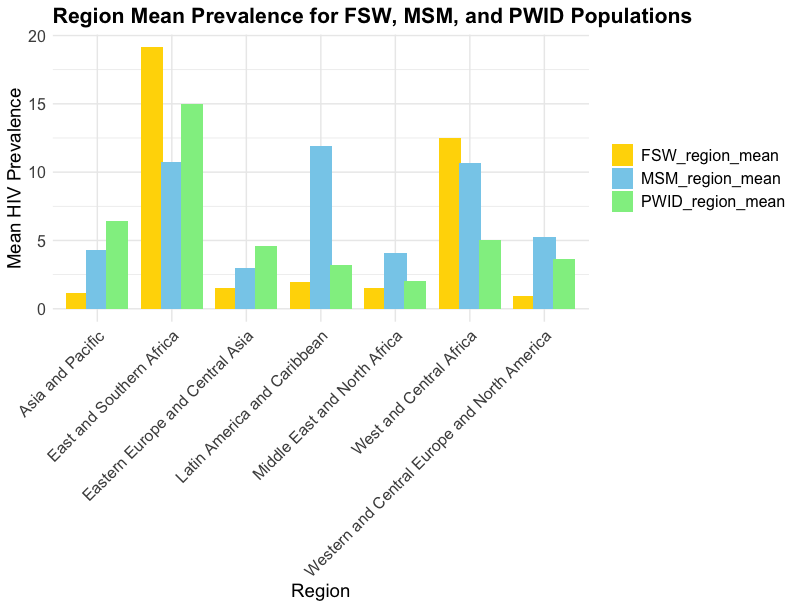}
    \caption{Regional mean HIV prevalence among MSM, PWID, and FSW populations across the seven UNAIDS-defined regions. The substantial differences across both regions and populations support the use of a region-specific mean structure in our model.}
    \label{fig:Region_Mean_dist}
\end{figure}

\noindent \textbf{Country Heterogeneity}: We consider two levels of country heterogeneity: region-level differences are modeled through region- and population-specific fixed effects, $\mu_K^{R(i)}$s, where $R(i)$ denotes the region that contains the country $i$.
The countries are grouped into seven UNAIDS-defined regions: Eastern and Southern Africa, Western and Central Africa, Middle East and North Africa, Asia and the Pacific, Eastern Europe and Central Asia, Western and Central Europe and North America, and Latin America and the Caribbean. Figure~\ref{fig:Region_Mean_dist} displays the regional mean HIV prevalence values. There are substantial differences between populations and regions. For example, HIV prevalence among FSW is markedly higher in African regions compared to Asia and the Pacific or Western and Central Europe and North America. Therefore, we assume that each region has its own region-specific mean for each key population. 

Within regions, country and population specific deviations from fixed effects $\mu_K^{R(i)}$ are captured by random effects $b_{ik}$, independently distributed as $b_{ik} \sim \mathcal{N}(0, \tau_k)$, where the variance $\tau_k$ is population-specific, implying all countries within a given population share the same degree of variability around the regional mean. This leads to the following linear mixed effects model:
\begin{equation}
\label{eq:LMM}
Y^{(i)}_{k,\,t} = \mu_k^{R(i)} + b_{ik} + \epsilon_{k,t}^{(i)},
\end{equation}

Instead of estimating the individual $b_{ik}$, we estimate $\tau_k$ and the covariance of $\epsilon_{k,t}^{(i)}$, so that the model is applicable to a data-sparse setting while preserving the flexibility to reflect different levels of heterogeneity. 

\noindent \textbf{Temporal Dependence}: For a given country \( i \) and key population \( k \), we further assume a first-order Markov structure over time: each year's prevalence is conditionally independent of all non-adjacent years given its immediate neighbors. 
For each key population, diagonal elements ($s$) represent the precision of individual years, and sub-diagonal elements ($\gamma$) capture correlations between adjacent years: ($s_1,\ \gamma_1$) for MSM, ($s_2$, $\gamma_2$) for FSW, and ($s_3$, $\gamma_3$) for PWID. The structure is illustrated below with an example precision matrix for MSM in country $i$ between 2011 and 2021, all unspecified entries are $0$s:
\begin{equation}
\label{eq:temp_Q}
Q_{\mathrm{MSM}}^{(i)}=\left[\begin{array}{ccccc}
s_{1} & \gamma_{1} & & & \\
\gamma_{1} & s_{1} & \gamma_{1} & & \\
& \gamma_{1} & s_{1} & \ddots & \\
& & \ddots & \ddots & \gamma_{1} \\
& & & \gamma_{1} & s_{1}
\end{array}\right]_{11 \times 11}
\end{equation}
By restricting the time dependency to adjacent years, the model captures short-term trends while maintaining computational efficiency.

\noindent \textbf{Cross-Population Dependence}:
We assume that prevalence rates between different key populations within the same year are dependent, reflecting shared year-specific factors such as policy changes, access to healthcare services, or HIV intervention programs. In contrast, prevalence rates for different populations across different years are assumed to be conditionally independent. For example, if the HIV prevalence rate is observed among PWID but missing for FSW in 2015, information from PWID in the same year may help infer that missing value, but data from PWID in other years would not provide additional information. 
The precision parameter $\rho_{k k'}$ describes the conditional dependence between the populations $k$ and $k'$. For example, the precision matrix between MSM and FSW in country $i$ is (all unspecified entries are 0):
\begin{equation}
\label{eq:cross_Q}
Q_{\text {MSM-FSW }}^{(i)}=\left[\begin{array}{lllll}
\rho_{12} & & & & \\
& \rho_{12} & & & \\
& & \rho_{12} & & \\
& & & \ddots & \\
& & & & \rho_{12}
\end{array}\right]_{11 \times 11}
\end{equation}

Similar precision matrices are defined for MSM-PWID and FSW-PWID using the parameters $\rho_{13}$ and $\rho_{23}$. We define a matrix $Q$ that incorporates temporal and cross-population precisions as follows:
\begin{equation}
\label{eq:combined_Q}
Q =
\begin{pmatrix}
Q_{\text{MSM}}^{(1)} & Q_{\text{MSM-FSW}}^{(1)} & Q_{\text{MSM-PWID}}^{(1)} & 0 & 0 & 0 & \cdots \\
Q_{\text{MSM-FSW}}^{(1)} & Q_{\text{FSW}}^{(1)} & Q_{\text{FSW-PWID}}^{(1)} & 0 & 0 & 0 & \cdots \\
Q_{\text{MSM-PWID}}^{(1)} & Q_{\text{FSW-PWID}}^{(1)} & Q_{\text{PWID}}^{(1)} & 0 & 0 & 0 & \cdots \\
0 & 0 & 0 & Q_{\text{MSM}}^{(2)} & Q_{\text{MSM-FSW}}^{(2)} & Q_{\text{MSM-PWID}}^{(2)} & \cdots \\
0 & 0 & 0 & Q_{\text{MSM-FSW}}^{(2)} & Q_{\text{FSW}}^{(2)} & Q_{\text{FSW-PWID}}^{(2)} & \cdots \\
0 & 0 & 0 & Q_{\text{MSM-PWID}}^{(2)} & Q_{\text{FSW-PWID}}^{(2)} & Q_{\text{PWID}}^{(2)} & \cdots \\
\vdots & \vdots & \vdots & \vdots & \vdots & \vdots & \ddots & \vdots 
\end{pmatrix}
\end{equation}

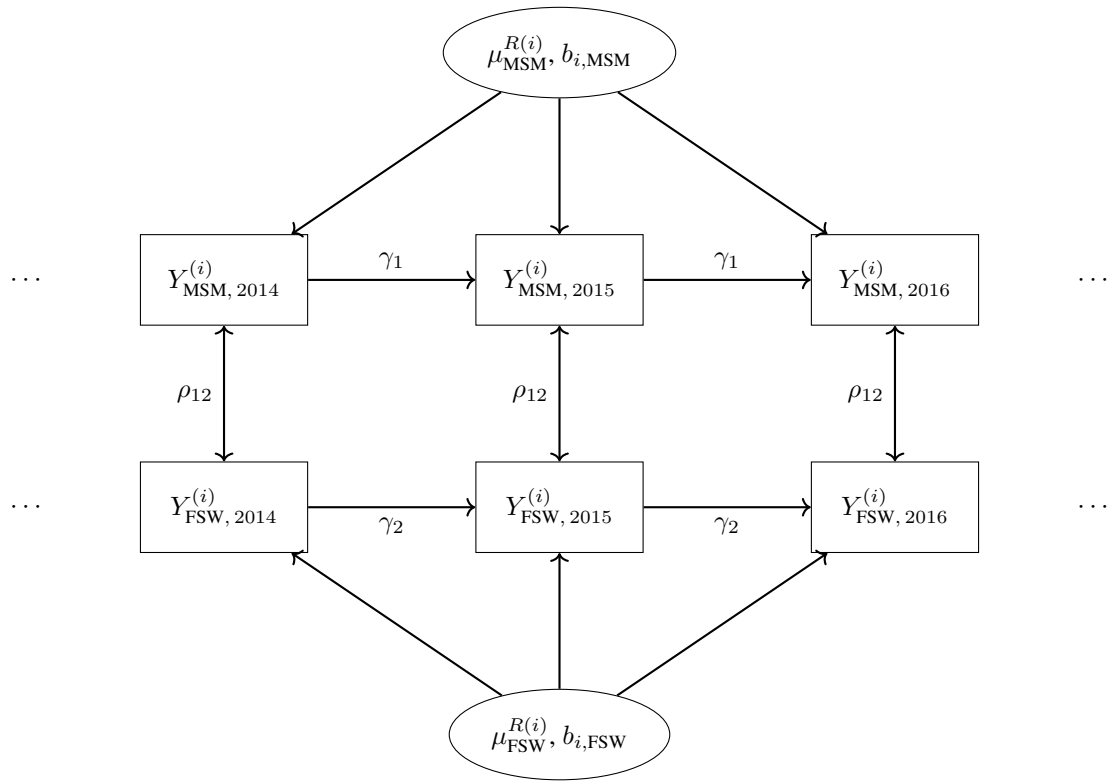
\begin{figure}[htbp!]
\centering
\begin{tikzpicture}[
  node distance=1.8cm and 2.2cm,
  every node/.style={font=\footnotesize},
  box/.style={draw, rectangle, minimum width=2.2cm, minimum height=1.2cm},
  oval/.style={draw, shape=ellipse, minimum width=2.5cm, minimum height=1.2cm},
  arrow/.style={->, thick},
  doublearrow/.style={<->, thick}
]

\node[box] (msm14) at (0,0) {$Y^{(i)}_{\text{MSM},\,2014}$};
\node[box, right=of msm14] (msm15) {$Y^{(i)}_{\text{MSM},\,2015}$};
\node[box, right=of msm15] (msm16) {$Y^{(i)}_{\text{MSM},\,2016}$};
\node[oval, above=of msm15] (tau_msm) {$\mu_{\text{MSM}}^{R(i)}$, $b_{i,\text{MSM}}$};

\node[box, below=of msm14] (fsw14) {$Y^{(i)}_{\text{FSW},\,2014}$};
\node[box, below=of msm15] (fsw15) {$Y^{(i)}_{\text{FSW},\,2015}$};
\node[box, below=of msm16] (fsw16) {$Y^{(i)}_{\text{FSW},\,2016}$};
\node[oval, below=of fsw15] (tau_fsw) {$\mu_{\text{FSW}}^{R(i)}$, $b_{i,\text{FSW}}$};

\draw[arrow] (msm14) -- (msm15) node[midway, above] {$\gamma_1$};
\draw[arrow] (msm15) -- (msm16) node[midway, above] {$\gamma_1$};

\draw[arrow] (fsw14) -- (fsw15) node[midway, below] {$\gamma_2$};
\draw[arrow] (fsw15) -- (fsw16) node[midway, below] {$\gamma_2$};

\draw[doublearrow] (msm14) -- (fsw14) node[midway, left] {$\rho_{12}$};
\draw[doublearrow] (msm15) -- (fsw15) node[midway, left] {$\rho_{12}$};
\draw[doublearrow] (msm16) -- (fsw16) node[midway, left] {$\rho_{12}$};

\draw[arrow] (tau_msm) -- (msm14);
\draw[arrow] (tau_msm) -- (msm15);
\draw[arrow] (tau_msm) -- (msm16);

\draw[arrow] (tau_fsw) -- (fsw14);
\draw[arrow] (tau_fsw) -- (fsw15);
\draw[arrow] (tau_fsw) -- (fsw16);

\node at ([xshift=-1.5cm]msm14.west) {$\cdots$};
\node at ([xshift=1.5cm]msm16.east) {$\cdots$};

\node at ([xshift=-1.5cm]fsw14.west) {$\cdots$};
\node at ([xshift=1.5cm]fsw16.east) {$\cdots$};

\end{tikzpicture}
\caption{Graphical representation of the dependence structure for two key populations (MSM and FSW) within a single country over three consecutive years (2014–2016). 
The model incorporates the region- and population specific mean ($\mu_{\text{MSM}}^{R(i)}$ and $\mu_{\text{FSW}}^{R(i)}$), the country-specific random effect ($b_{ik}$), additional temporal dependencies within each population ($\gamma_1$, $\gamma_2$), and additional cross-population correlations between MSM and FSW ($\rho_{12}$). The arrows represent the structured dependencies encoded in the model.}
\label{fig:model_structure}
\end{figure}

\noindent \textbf{Covariance Matrix}: Finally, we present the full cross-population hierarchical model for HIV prevalence estimation, which integrates spatial, temporal, and cross-population dependencies into a unified statistical framework:
\begin{equation}
\label{eq:final_model}
\boldsymbol{Y} \sim \mathcal{N}(\boldsymbol{\mu},\ Q^{-1} +\ \Omega).
\end{equation}
\begin{itemize}
    \item \(\boldsymbol{Y} = [Y_{MSM}^{(1)}, Y_{FSW}^{(1)}, Y_{PWID}^{(1)}, Y_{MSM}^{(2)}, Y_{FSW}^{(2)}, Y_{PWID}^{(2)}\ldots, Y_{PWID}^{(N)}]^\top\) is the vector of log transformed HIV prevalence values for all countries, key populations, and years, where $N$ is the number of countries. For country \( i \) and population \( k \in \{\text{MSM}, \text{FSW}, \text{PWID}\} \), the subvector \( Y^{(i)}_k \in \mathbb{R}^{11} \) corresponds to the annual HIV prevalence rates from 2011 to 2021.
    \item $\boldsymbol{\mu}$ is the corresponding mean vector as defined in Equation~\ref{eq:LMM}. 
    \item \(Q^{-1}\) denotes the inverse of the precision matrix in Equation~\ref{eq:combined_Q}, encoding correlations over years and between key populations within each country. 
    \item $\Omega = \textup{diag}(T^{(1)}_{MSM}, T^{(1)}_{FSW}, T^{(1)}_{PWID}, T^{(2)}_{MSM}, \ldots, T^{(N)}_{FSW}, T^{(N)}_{PWID})$ is the block-diagonal covariance matrix of country-specific random effects, where $T^{(i)}_k$ is a 11$\times$11 constant matrix with all elements equal to $\tau_k$. 
\end{itemize}
This formulation allows us to jointly model prevalence across populations while preserving local dependencies, ensuring both flexibility and computational efficiency. We provide a graphical representation to summarize the dependence structure in our model in Figure~\ref{fig:model_structure}. Next, we will describe Bayesian inference for model parameters.


\subsection{Posterior inference for observed and unobserved outcomes}
\label{sec:bayes}

The model is implemented in \textit{Stan} using a Bayesian framework that incorporates prior knowledge, explicitly handles uncertainty, and supports probabilistic parameter estimation. To stabilize inference under data sparsity, we apply a Laplace(0, 0.1) prior to all parameters, encouraging shrinkage of weak effects towards zero and improving convergence. Sensitivity analysis (see Appendix) showed that results were robust under a less informative Laplace(0, 0.5) prior; however, non-sparse priors such as Normal(0, 0.1) produced unstable posteriors.

To obtain posterior draws of HIV prevalence for all countries, populations, and years, we distinguish between observed values $Y_{\text{obs}}$ and unobserved (missing) values $Y_{\text{miss}}$. Conditional on model parameters $(\mu, \Sigma)$, the joint distribution of $Y$ follows a multivariate normal law. Reordering the covariance matrix to align with the observed--missing partition yields
\[
\Sigma =
\begin{pmatrix}
\Sigma_{\text{obs,obs}} & \Sigma_{\text{obs,miss}} \\
\Sigma_{\text{miss,obs}} & \Sigma_{\text{miss,miss}}
\end{pmatrix},
\]
where $\Sigma_{\text{obs,obs}}$ denotes the covariance among observed entries, $\Sigma_{\text{miss,miss}}$ the covariance among missing entries, and $\Sigma_{\text{miss,obs}}$ the cross-covariance. 

\paragraph{Posterior predictive distribution for unobserved values.}  
The conditional distribution of $Y_{\text{miss}}$ given $Y_{\text{obs}}$ follows a multivariate normal distribution:
\begin{equation}
\label{eq:conditional}
Y_{\text{miss}} \mid Y_{\text{obs}}, \mu, \Sigma \;\sim\; 
\mathcal{N}\!\left( 
\mu_{\text{miss}} +
\Sigma_{\text{miss,obs}} \Sigma_{\text{obs,obs}}^{-1}(Y_{\text{obs}}-\mu_{\text{obs}}),\;
\Sigma_{\text{miss,miss}} - \Sigma_{\text{miss,obs}} \Sigma_{\text{obs,obs}}^{-1} \Sigma_{\text{obs,miss}}
\right).
\end{equation}
For each posterior draw of $(\mu,\Sigma)$, we sample from the above distribution to impute the missing entries. This approach propagates both parameter uncertainty and data uncertainty.

\paragraph{Posterior reconstruction of observed values.}  
Observed entries $Y_{\text{obs}}$ are also treated as random variables within the hierarchical model, combining country-level random effects $b_{ik}$ and structured errors $\epsilon_{kt}^{(i)}$. To obtain posterior draws of $Y_{\text{obs}}$, we employ an alternating Gibbs sampler:

\begin{tcolorbox}[title=Gibbs Sampling Algorithm for \texorpdfstring{$b_{ik}$}{bik} and \texorpdfstring{$\epsilon_{kt}^{(i)}$}{e_kti}]
\begin{enumerate}
  \item \textbf{Update random effects $b_{ik}$.}  
  For each country $i$ and population $k$, conditional on current $\epsilon_{kt}^{(i)}$, draw
  \[
  b_{ik} \sim \mathcal{N}\left(
  \frac{1}{\tfrac{1}{\tau_k} + n_{ik}} \sum_{t \in \mathcal{T}_{ik}} 
  \left( Y_{kt}^{(i)} - \mu_k^{R(i)} - \epsilon_{kt}^{(i)} \right),\;
  \left( \tfrac{1}{\tau_k} + n_{ik} \right)^{-1}
  \right),
  \]
  where $\mathcal{T}_{ik}$ is the set of observed years for $(i,k)$ and $n_{ik}=|\mathcal{T}_{ik}|$.
  
  \item \textbf{Update errors $\epsilon_{kt}^{(i)}$.}  
  The error vector $\boldsymbol{\epsilon}$ follows a multivariate normal prior $\mathcal{N}(\mathbf{0}, Q^{-1})$, with precision matrix $Q$ determined by the spatio–temporal Gaussian Markov random field. For index $j$ corresponding to $(k,t,i)$, draw
  \[
  \epsilon_j \mid \boldsymbol{\epsilon}_{-j} \sim \mathcal{N}\left(
  -\frac{1}{Q_{jj}} \sum_{l \ne j} Q_{jl} \epsilon_l,\; Q_{jj}^{-1}
  \right).
  \]

  \item \textbf{Reconstruct $Y_{\text{obs}}$.}  
  With updated $b_{ik}$ and $\epsilon_{kt}^{(i)}$, generate posterior draws of the observed outcomes via
  \[
  Y_{kt}^{(i)} = \mu_k^{R(i)} + b_{ik} + \epsilon_{kt}^{(i)},
  \quad (k,t,i) \in \text{observed set}.
  \]
\end{enumerate}
\end{tcolorbox}

\paragraph{Posterior summaries.}  
We repeat this procedure for 12,000 posterior draws of $(\mu,\Sigma)$, yielding full posterior samples of $Y$ (both observed and missing) for all countries, populations, and years. This enables us to compute posterior summaries and credible intervals for temporal contrasts, such as the change between 2011 and 2021, for progress evaluation purposes.


\section{Results}
\label{sec:result}
We use the regional median HIV prevalence for each key population as the baseline estimate for unobserved data (the current practice in global reporting).
We present the numerical results of the baseline approach, the proposed model, and its simplified variants. The computational experiments were conducted on the Penn State ROAR high-performance computing cloud, using 12 cores, each running one of 12 parallel sampling chains with 8 GB of memory per core. Generating 12,000 samples took approximately 38 hours and 40 minutes.



We evaluate the prediction accuracy using 5-fold cross-validation. All metrics are computed on the log-transformed HIV prevalence scale, consistent with the model's outcome specification. The proposed model achieves a cross-validated mean squared error (MSE) of 0.47, while the baseline’s cross-validated MSE is 1.33. This 65\% reduction in error underscores the benefits of modeling regional, temporal, and cross-population dependencies in the model. We calculate the empirical coverage of the 95\% posterior predictive intervals for the test data. The proposed model has a coverage of 93.39\%, indicating that the posterior intervals are well calibrated, ensuring the reliability of estimates for decision-making in data-sparse settings. 

\begin{figure}[htbp!]
    \centering
    \includegraphics[width=1\textwidth]{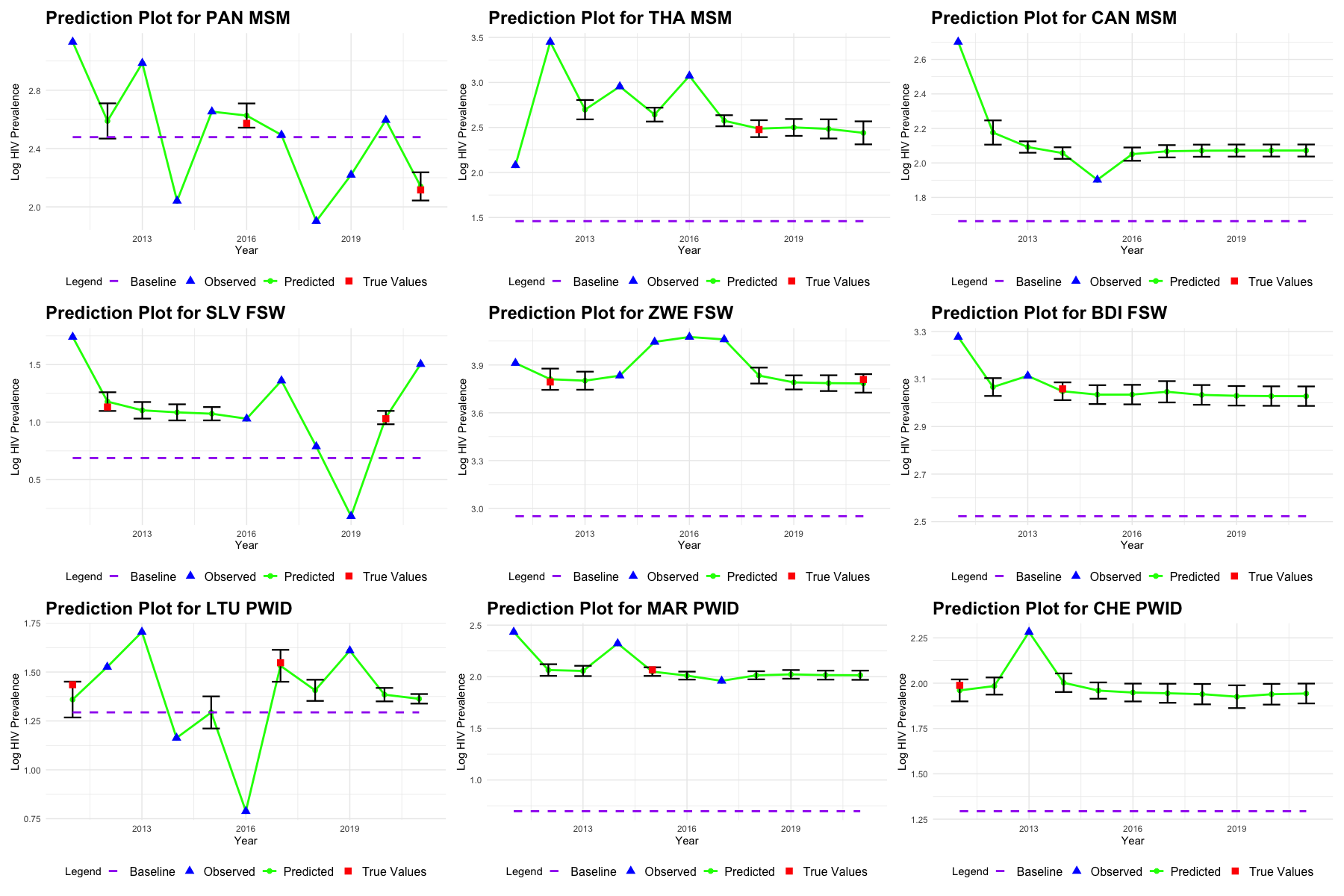}
    \caption{Prediction plots for each key population (MSM, FSW, and PWID) across three countries representing different levels of data availability: sparse, moderate, and dense (from left to right). The plots compare data points withheld for testing (red squares), baseline estimates (dashed purple line), model prediction (green lines). Blue triangles are data points used in fitting and the error bars show 95\% posterior predictive intervals. PAN = Panama, THA = Thailand, CAN = Canada, SLV = El Salvador, ZWE = Zimbabwe, BDI = Burundi, LTU = Lithuania, MAR = Morocco, CHE = Switzerland.}
    \label{fig:country_result}
\end{figure}

For each key population, we chose three representative countries with varying data availability ('dense', 'moderate' and 'sparse') to visualize the estimated prevalence of HIV. In particular, we selected Panama (dense), Thailand (moderate), and Canada (sparse) for MSM; El Salvador (dense), Zimbabwe (moderate), and Burundi (sparse) for FSW; and Lithuania (dense), Morocco (moderate), and Switzerland (sparse) for PWID. Figure \ref{fig:country_result} shows the results for the nine representative countries. Our proposed model consistently outperforms the baseline across all cases, demonstrating its robustness in handling different levels of data sparsity.



\begin{figure}[htbp!]
    \centering
    \includegraphics[width=0.8\textwidth]{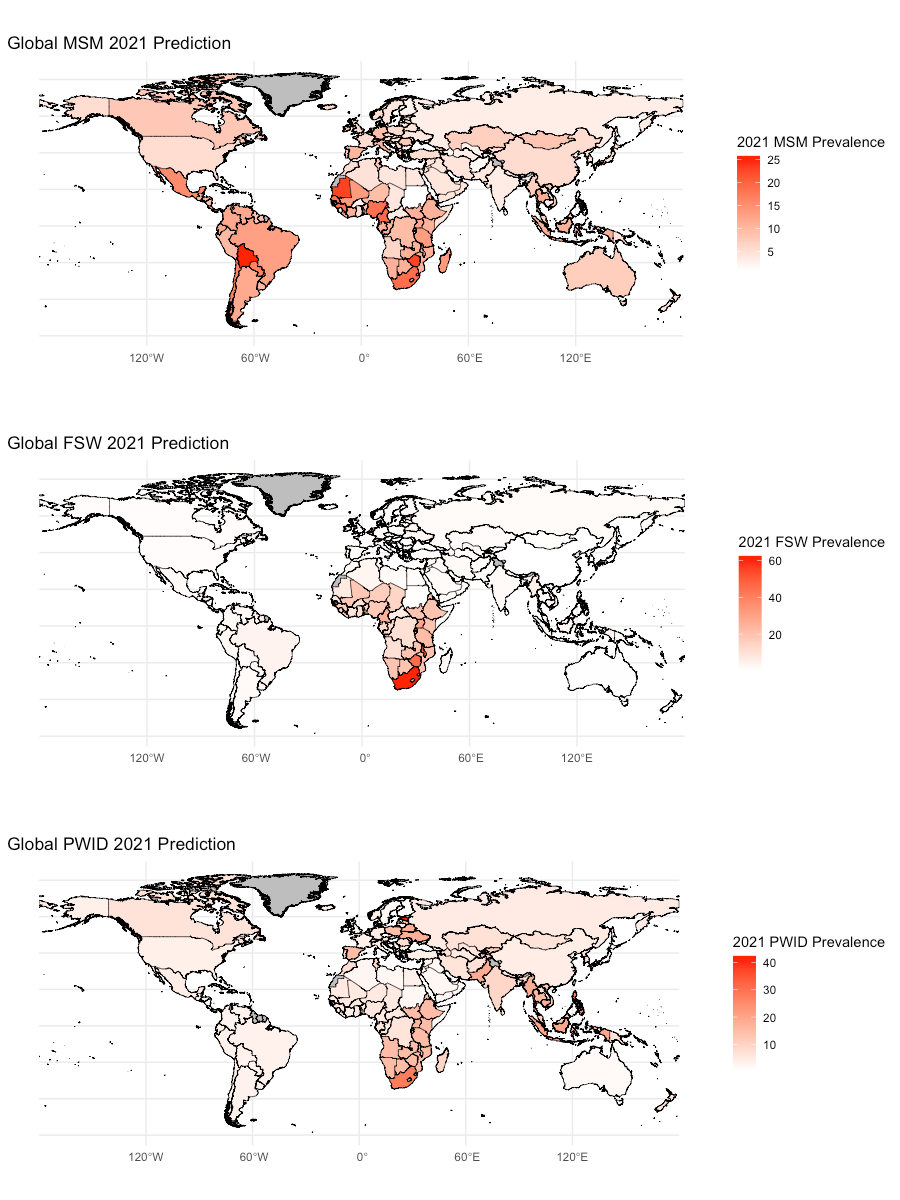}
    \caption{Estimated HIV prevalence in 2021 for three key populations: MSM (top), FSW (middle), and PWID (bottom). Color intensity reflects prevalence on the original scale (after exponentiation, i.e. exp($Y$)). The maps highlight strong geographic heterogeneity, with FSW showing the highest predicted prevalence levels in sub-Saharan Africa, MSM in parts of Latin America and Africa, and PWID in Southeast Asia and select Eastern European countries.}
    \label{fig:global_2021}
\end{figure}

We map the estimated HIV prevalence in 2021 for all countries to visualize the geographic heterogeneity (Figure~\ref{fig:global_2021}). The HIV prevalence rates among three key populations are notably high in sub-Saharan Africa. Latin America presents a similar high HIV prevalence among MSM, while Southeast Asia and select parts of Eastern Europe show a high level of HIV prevalence among PWID.


\begin{figure}[h!]
    \centering
    \includegraphics[width=0.8\textwidth]{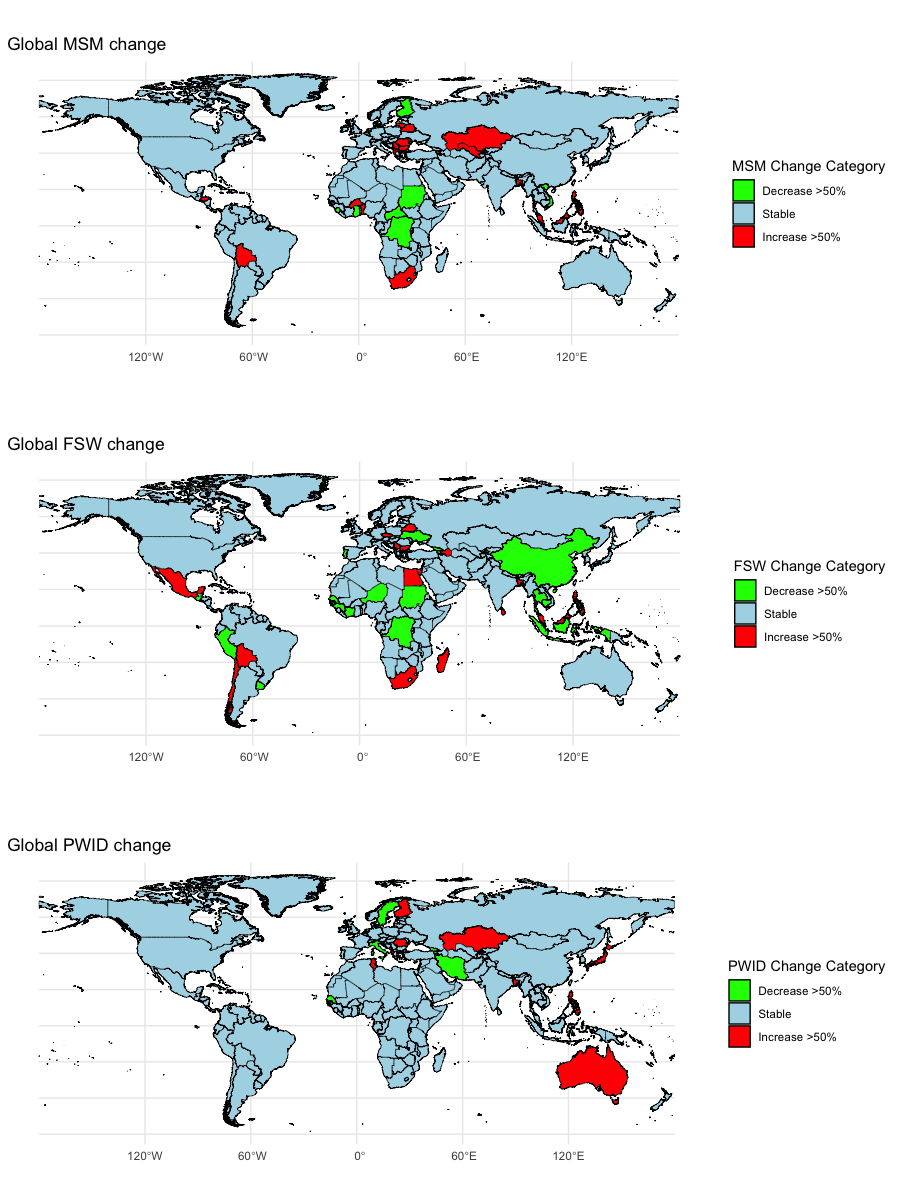}
    \caption{Global changes in HIV prevalence between 2011 and 2021 for MSM (top), FSW (middle), and PWID (bottom). Countries are colored by the direction and magnitude of change using $\Pr\!\big(ratio>1.5\big) > 0.95$ (increase) or $\Pr\!\big(ratio<0.5\big) > 0.95$ (decrease). Red indicates a statistically significant increase of more than 50\%, green indicates a significant decrease of more than 50\%, and blue indicates no significant change. These maps highlight regions experiencing substantial shifts in epidemic burden over the past decade.}
    \label{fig:change_map}
\end{figure}

To assess changes in HIV prevalence over time, we focused on the contrast between 2011 and 2021 for each key population and each country. For each country–population combination, we computed posterior draws of the log-difference 
\[
\Delta = Y_{2021} - Y_{2011},
\]
where $Y_t$ denotes the log prevalence in year $t$. To aid interpretation, we transformed these differences to the original prevalence scale by exponentiation:
\[
\text{Ratio} = \exp(\Delta) = \frac{\exp(Y_{2021})}{\exp(Y_{2011})}.
\]
Figure~\ref{fig:change_map} highlights countries showing substantial changes between 2011 and 2021, defined as at least a 50\% increase or decrease—supported by posterior probability $>0.95$.

We observe notable increases in HIV prevalence among MSM in parts of Eastern Europe, Central Asia, and South America, while significant declines are seen in several countries in sub-Saharan Africa. For FSW, countries such as South Africa and Ghana show marked decreases, while parts of South America and Southeast Asia exhibit rising trends. Among PWID, sharp increases are observed in Australia and Eastern Europe, whereas Western Europe and parts of Africa demonstrate substantial declines. These trends emphasize the evolving nature of the HIV epidemic and the importance of maintaining up-to-date surveillance and adaptive intervention strategies.

Table \ref{tab:corr} summarizes the posterior correlations between key populations (FSW, MSM, and PWID) and temporal correlations between years, with 95\% credible intervals. The diagonal entries indicate temporal correlations within each population, whereas the off-diagonal entries represent cross-population dependencies. The temporal dependencies are much stronger than the cross-population dependencies: 0.81 for MSM, 0.54 for FSW, and 0.68 for PWID. 
The small correlation between MSM and FSW (0.0026) possibly reflects the fact that the MSM and FSW populations do not overlap and have distinct social networks. The modest positive correlations for FSW-PWID (0.041) and MSM-PWID (0.035) imply some shared risk environments, and these pairs of populations could also overlap.
Although these cross-population correlations are not large in magnitude, they all differ significantly from zero and improve prediction accuracy, i.e., the cross-validated MSE increases when this component is removed. 

\begin{table}
\caption{\label{tab:corr}
Estimated posterior means and 95\% credible intervals for cross-population and temporal correlations. Symmetric values are omitted for simplicity. MSM = Men who have sex with men, FSW = Female sex workers, PWID = People who inject drugs.}
\begin{tabular}{c|lll}
\hline
\textbf{Populations} & \textit{MSM} & \textit{FSW} & \textit{PWID} \\ 
\hline
\textit{MSM} & 0.810 (0.808, 0.812) & 0.0026 (0.0018, 0.0033) & 0.035 (0.034, 0.036) \\
\textit{FSW} & - & 0.544 (0.542, 0.546) & 0.041 (0.040, 0.042) \\
\textit{PWID} & - & - & 0.683 (0.681,0.685) \\
\hline
\end{tabular}
\end{table}

Finally, we evaluate the contributions of the three key dependencies (country, temporal, and cross-population effects) to prediction accuracy by removing each dependence parameter from the model or equivalently fixing the parameter value at zero. 
Table \ref{tab:mse_results} confirms that the sampling algorithm converged well for all models, as evidenced by $\hat R \le 1.01$ and high effective sample sizes.
The statistic $\hat{R}$ assesses the convergence of MCMC, where $\hat{R}$ values close to 1 indicates well-mixed sampling chains. 
The effective sample size, $n_{\text{eff}}$, quantifies the number of independent draws from the posterior, adjusting for autocorrelation; values above 1,000 are generally considered sufficient for stable parameter estimation. 
Table \ref{tab:mse_results} also summarizes the cross-validated MSE for the full model, the reduced models, and the baseline model (regional median). The full model has the lowest MSE (0.47), outperforming all reduced models and the baseline model (1.33). 
Removing the temporal dependence has the most severe impact, causing the MSE to increase dramatically to 18.38. Excluding country-specific random effects also substantially degrades performance (MSE = 0.58). Removing the dependence between populations results in a marginal increase in MSE. 

\begin{table}
\caption{\label{tab:mse_results}
Convergence diagnosis ($\hat{R}$ 
and effective sample size - $n_{\text{eff}}$ with median, 
min and max values) and 
cross-validated mean squared errors (CV MSE) for the full model, 
reduced models (removing one effect at a time), and the baseline model (regional median).}
\begin{tabular}{|c|c|c|c|c|c|}
\hline
\textbf{Model} & \textbf{CV MSE} & $\hat{R}$ & \textbf{$n_{\text{eff}}$ - median} & \textbf{$n_{\text{eff}}$ - min} & 
\textbf{$n_{\text{eff}}$ - max} \\
\hline
Full Model & 0.4677 & 1.00 & 6298.5 & 4953 & 8594\\
No Cross-Pop Effect & 0.4971 & 1.00 & 6888 & 6006 & 9200 \\
No Country Effect & 0.5852 & 1.00 & 4337 & 3781 & 5890 \\
No Time Effect & 18.3846 & 1.00 & 6736 & 4242 & 7210\\
Baseline & 1.3306 & NA & NA & NA & NA \\
\hline
\end{tabular}
\end{table}

\section{Conclusion and Discussion}
\label{sec:Conclusion}

Key populations (female sex workers, men who have sex with men, and people who inject drugs) bear a disproportionate share of the HIV burden, yet surveillance for these groups is sparse, uneven across countries, and often summarized by simple regional medians. In this study, we develop a cross-population hierarchical model to estimate HIV prevalence rates among key populations, addressing challenges such as data sparsity and different levels of heterogeneity. 
Our global analysis produces globally comparable and uncertainty-quantified country-by-year prevalence estimates for these populations worldwide by borrowing strength across space, time, and populations. 

It reveals divergent trends in HIV prevalence among key populations between 2011 and 2021. Substantial declines among female sex workers (FSW) in sub-Saharan Africa suggest that intensified prevention and treatment efforts—including condom promotion, community outreach, and antiretroviral coverage—are yielding measurable gains. In contrast, the persistence or increase of prevalence among men who have sex with men (MSM) in Eastern Europe and Central Asia indicates ongoing challenges such as stigma, criminalization, and barriers to accessing services.


Our modeling framework contributes to the scientific toolkit by systematically bridging data gaps across space, time, and populations—including MSM, FSW, and people who inject drugs (PWID)—and propagating full posterior uncertainty. With a 65\% reduction in cross-validated mean squared error compared to baseline methods, we demonstrate both predictive reliability and analytical transparency. The methodological framework also has broader applicability beyond HIV, offering potential for use in other public health domains, such as tuberculosis or malaria, where there are similar challenges of sparse data and complex dependency structures. 

Nonetheless, limitations remain. We assume constant cross-population correlation parameters globally, given data sparsity; these average effects may mask regional differences. The correlation can potentially vary by region or country. If sufficient data become available, future models could explore region-specific cross-population dependencies to capture local epidemiological patterns more precisely.

We do not incorporate an adjacency-based spatial correlation; instead, we use UNAIDS-defined regions to capture primary geographic heterogeneity. This decision was motivated by data sparsity at the country level. In UNAIDS region classification, for example, instead of treating Africa as a single region, it partitions African countries into Eastern and Southern Africa, Western and Central Africa, and the Middle East and North Africa. During preliminary testing, we found that the conditional autoregressive model with adjacency-based dependence provided minimal improvement in predictive performance.

Our model treats all observed HIV prevalence values as equally reliable point estimates. However, in practice, data quality varies significantly between studies in terms of sample size, methodology, and potential measurement error. Due to the limited and inconsistent reporting of uncertainty measures across countries, we did not incorporate these variations into our model. Future extensions could account for such heterogeneity by explicitly modeling measurement error or applying weights based on survey quality, where such information is available.

Finally, it is also important to note that our modeling approach is not a substitute for comprehensive data collection. In fact, its accuracy and utility would be significantly enhanced by improved data quality and expanded coverage, particularly for underrepresented populations and regions; rather, it should be viewed as a complementary tool that maximizes the value of existing data.



\section*{Acknowledgements}
Jiahao Zhang and Le Bao's work was partially supported by NIH / NIAID R01AI136664 and R01AI170249.

\section*{Supplementary Materials}
\addcontentsline{toc}{section}{Appendices}
\renewcommand{\thesubsection}{\Alph{subsection}}

\subsection{List of Notations}
\textcolor{black}{\bf The list of primal variables and indices:}
\begin{itemize}
\item $N$: the total number of countries.
\item $\boldsymbol{Y}$: the vector of log transformed HIV prevalence values, where each entry corresponds to a specific country, year, and key population. 
\item $Y^{(i)}_{k,\,t}$: entry of $\boldsymbol{Y}$, where \( i \in \{1,2,...,N\}\) indexes countries, \( k \in \{\text{MSM}, \text{FSW}, \text{PWID}\} \) denotes key population, and \( t \in \{2011, \ldots, 2021\} \) indicates year.
\item $\boldsymbol{\mu}$: the mean vector.
\item $\Sigma$: a structured covariance matrix encoding dependencies across time, populations, and countries.
\item $s_1, s_2, s_3$: diagonal elements of the temporal effect precision matrix block of MSM, FSW, and PWID correspondingly.
\item $\gamma_1, \gamma_2, \gamma_3$: sub-diagonal elements of the temporal effect precision matrix of MSM, FSW, and PWID for capturing correlations between adjacent years.
\item $\rho_{12}$, $\rho_{13}$, $\rho_{23}$: diagonal elements of the cross-population precision block among MSM, FSW, and PWID for capturing correlations between populations.
\item $Q_{\mathrm{k}}^{(i)}$: temporal effect precision matrix block for population $k$ and country $i$.
\item $Q_{\substack{k_1\text{-}k_2}}^{(i)}$: cross-population effect precision matrix block between population $k_1$ and $k_2$ for country $i$.
\item $\tau_1, \tau_2, \tau_3$: within-region variance correspond to MSM, FSW, and PWID.
\item $T^{(i)}_{k}$: random effect matrix block for population $k$ and country $i$ filled by $\tau_k$.
\item $\Omega$: integrated random effect matrix where diagonals are $T^{(i)}_{k}$.
\item $Q$: integrated precision matrix combined temporal effect precision blocks and cross-population effect precision blocks.
\end{itemize}

\subsection{Prior Sensitivity Analysis}
We conducted a prior sensitivity analysis to evaluate the influence of our chosen prior on model estimates, particularly for cross-population correlation parameters. The main model uses a Laplace(0, 0.1) prior on all parameters to encourage sparsity and stabilize estimation in the presence of extreme data sparsity. To test robustness, we fit the model under the following alternative priors:

\begin{itemize}
    \item \textbf{Laplace(0, 0.5)}: a less aggressive sparsity-inducing prior
    \item \textbf{Normal(0, 0.1)}: a weakly informative, non-sparse prior
\end{itemize}

Each model was run with the same data, initial values, and Stan settings. Posterior summaries of cross-population correlations are shown in Table~\ref{tab:prior_sensitivity}.

\begin{center}
\textbf{Table. Sensitivity of posterior estimates and CV MSE under different prior settings.}
\label{tab:prior_sensitivity}
\vspace{0.5em}

\begin{tabular}{lcccc}
\hline
\textbf{Prior} & \textbf{MSM--FSW} & \textbf{MSM--PWID} & \textbf{FSW--PWID} & \textbf{CV MSE} \\
\hline
Laplace(0, 0.1) & 0.0026 & 0.035 & 0.041 & 0.4804 \\
Laplace(0, 0.5) & 0.0031 & 0.036 & 0.042 & 0.4832 \\
Normal(0, 0.1)  & 0.0018--0.011$^{*}$ & 0.031--0.052$^{*}$ & 0.032--0.061$^{*}$ & 0.496--0.810$^{*}$ \\
\hline
\end{tabular}

\vspace{0.5em}

{\footnotesize \textit{Note.} (*) means that the Normal(0, 0.1) prior exhibited noticeably higher variation in posterior estimates across multiple sampling runs, indicating reduced stability and convergence challenges. In contrast, Laplace priors yielded stable and consistent results, Laplace (0, 0.5) gives similar estimated results but with higher CV MSE.}
\end{center}

The Laplace(0, 0.1) prior was selected to ensure computational stability and regularize weak effects, especially in settings with sparse or non-overlapping data across populations. While a less aggressive Laplace(0, 0.5) prior led to similar results, using a non-sparse prior like Normal(0, 0.1) caused unstable inference and inconsistent parameter estimates across repeated fits, despite unchanged data and settings. These results support the robustness of our modeling conclusions and justify the use of a strongly regularizing prior in the main analysis.

\subsection{Data and Code}
\label{sec:codes}
Data and code are available as the online supplementary materials.

\bibliographystyle{rss}

\end{document}